\documentclass[a4paper,reqno]{article}
\usepackage{amsmath,amssymb,amstext}
\usepackage[body={15cm,20.5cm}]{geometry}
\usepackage{url}
\usepackage{graphicx}
\graphicspath{{growth-figs/}}

\begin{document}

\begin{center}

{\Large On Exact Multi-fractal Spectrum of the Whole-Plane SLE}\\

\vspace{5mm}

Igor \textsc{Loutsenko}

\vspace{3mm}

Universit\'e Paris Diderot\\
e-mail: loutsenko@math.jussieu.fr\\[1mm]

\vspace{3mm}

Oksana \textsc{Yermolayeva}

\vspace{3mm}

Institut de Physique Th\'eorique, Saclay\\
e-mail: oksana.yermolayeva@cea.fr\\[1mm]


\vspace{3mm}

November 23, 2012\\

\vspace{5mm}

Abstract

\end{center}

\begin{quote}
We consider the whole-plane
Shramm-Loewner evolution. Using exact solutions of fundamental equations
for moments of derivative of conformal mappings we determine its average integral means beta-spectrum.
\end{quote}

\section{Introduction}

Recently, the area of two-dimensional critical phenomena has enjoyed a breakthrough due to a radical development by O.~Schramm in stochastic Loewner evolution, as an approach to description of boundaries of critical clusters (see for a review e.g. \cite{G} and \cite{GL2}). The most  important characteristics of such boundaries are their multifractal spectra.  In this article we give exact description of the average $\beta-$spectrum for unbounded conformal mappings of the whole-plane SLE$_\kappa$.

An initial motivation for the present work was study of the coefficient problem for injective conformal mappings from the interior of the unit circle to the plane determined by the whole-plane Stochastic Loewner Evolution (SLE) flows \cite{DNNZ}, \cite{L}.
Investigation of the asymptotic behavior of the moments of the Taylor coefficients of such mappings led us to finding the exact multi-fractal spectrum of unbounded conformal mappings related to the whole-plane SLE-flows.

In this section we introduce notations and describe main results.

Radial stochastic Schramm-Loewner Evolution with parameter $\kappa$ (see e.g. \cite{C}, \cite{GL}, \cite{GL2})  describes dynamics of a slit domain in the $z$-plane that can be represented by growth of a random planar curve $\Gamma=\Gamma(t)$ starting from a point on a unit circle $|z|=1$ at $t=0$. The principal idea of the theory of SLE is that the growth of the curve can be given by the time dependent conformal mapping $z=F(w,t)$ which satisfies the Loewner equation with stochastic driving:
$$
\frac{\partial F(w,t)}{\partial t}=w\frac{\partial F(w,t)}{\partial w}\frac{w+e^{\mathrm{i} \tilde B(t)}}{w-e^{\mathrm{i} \tilde B(t)}}, \quad  t\ge 0, \quad F(w,0)=w,
$$
where $\tilde B(t)$ denotes the Brownian motion with ``temperature'' $\kappa$:
$$
\langle (\tilde B(t)-\tilde B(t'))^2 \rangle =\kappa |t-t'|.
$$
Everywhere through the article $\langle\,\,\rangle$ denotes expectation (ensemble averaging).

One can consider either the ``exterior'' problem, where the complement of the unit disc $\mathbb{D}_+=\{w: |w|>1\}$ in the $w$-plane is mapped by $F=F_+$ to the ``exterior" slit domain $\mathbb{D}_+\backslash\Gamma(t)$ in the $z$-plane (i.e. the curve starting from the unit circle is growing in the exterior of the unit disc)
$$
F_+(w,t)=e^t\left(w+\sum_{i=0}^\infty\frac{F_i^+(t)}{w^i}\right), \quad |w|>1,
$$
or the ``interior'' problem, where the curve starting from the unit circle is growing in the interior of the unit disc. In this problem the unit disc in the $w$-plane $\mathbb{D}_-=\{w: |w|<1\}$ is mapped by $F=F_-$ to the ``interior" slit domain $\mathbb{D}_-\backslash\Gamma(t)$ in the $z$-plane
$$
F_-(w,t)=e^{-t}\left(w+\sum_{i=2}^\infty F_i^-(t)w^i\right), \quad |w|<1.
$$
Since SLE$_\kappa$ is a conformally invariant stochastic process \cite{C}, \cite{G}, \cite{GL}, the exterior and interior problems are related by the inversion $F_-(w,t)=1/F_+(1/w,t)$.

In the present article we study the whole-plane SLE, which is an infinite-time limit of the radial SLE: There are two versions of the whole-plane SLE$_\kappa$:
\begin{equation}
\frac{\partial \mathcal{F}_\pm(w,t)}{\partial t}=\pm w\frac{\partial \mathcal{F}_\pm(w,t)}{\partial w}\frac{w+e^{\mathrm{i} B(t)}}{w-e^{\mathrm{i} B (t)}}, \quad \mathcal{F}_\pm(w,t)=e^{t}\left(w+\sum_{i=1\mp 1}^\infty \mathcal{F}_i^\pm(t)w^{\mp i}\right) .
\label{WholeSLE_fixed}
\end{equation}

The first version is the "interior" whole plane SLE$_\kappa$ which can be viewed as the limit of the interior problem
\begin{equation}
\mathcal{F}_-(w,t)=\lim_{T\to\infty}e^TF_-(w,T-t), \quad |w|<1, \quad B(t)=\tilde B(T-t)
\label{T_limit_SLE_int}
\end{equation}
describing the growth process in an infinite slit domain by ``erasing'' in time a curve/slit that starts at some point on the plane and goes to infinity.

The second version of the whole-plane SLE$_\kappa$ can be viewed as the limit of the exterior problem
\begin{equation}
\mathcal{F}_+(w,t)=\lim_{T\to\infty}e^{-T}F_+(w,T+t), \quad |w|>1,  \quad B(t)=\tilde B(T+t)
\label{T_limit_SLE_ext}
\end{equation}

In the present article we find explicit expression for average $\beta(q)$-spectrum for the unbounded-map (interior) whole-plane SLE$_\kappa$. In more details, the $\beta$-spectrum quantitatively describes subsets of the domain boundary, where certain scaling laws apply to harmonic measure. This describes singularities of derivatives of conformal mappings related to distribution of the ``wedge angles" along the curve (see e.g. \cite{BS}, \cite{BS2}, \cite{D} and for general introduction to multi-fractal analysis see e.g. \cite{HJKPS}). Below we study the integral means $\beta(q)$ spectrum which is defined through the moments of the derivatives at $|w|\to 1$:
\begin{equation}
\beta(q)=\lim_{\epsilon\to0+}\frac{\log \int_0^{2\pi}\langle|\mathcal{F}_\pm'\left(e^{\pm\epsilon+\mathrm{i}\varphi},t\right)|^q\rangle d\varphi}{-\log \epsilon},
\label{beta_spectrum}
\end{equation}
where the prime denotes the $w$-derivative and $\pm$ stand for the exterior or interior cases respectively. Below we present arguments  related to the following result: The average integral means $\beta$-spectrum of the interior whole-plane SLE$_\kappa$ equals
\begin{equation}
\beta=\left\{\begin{array}{ll}
\kappa\frac{\gamma(q,\kappa)^2}{2}-2\gamma(q,\kappa)-1 , &  \quad q\le -1-\frac{3\kappa}{8} \\
\kappa\frac{\gamma(q,\kappa)^2}{2}, & \quad -1-\frac{3\kappa}{8} \le q \le Q(\kappa) \\
3q-\frac{1}{2}-\frac{1}{2}\sqrt{1+2q\kappa}, & \quad  q \ge Q(\kappa)
\end{array}\right.
\label{results}
\end{equation}
where
\begin{equation}
\gamma(q,\kappa)=\frac{\kappa+4-\sqrt{(\kappa+4)^2-8q\kappa}}{2\kappa}
\label{gamma}
\end{equation}
and
\begin{equation}
Q(\kappa)=\frac{\kappa^2+8\kappa+12-2\sqrt{2\kappa^2+16\kappa+36}}{16\kappa}
\label{Q}
\end{equation}
The arguments are based on the infinite number of one-parametric families of exact solutions for moments of derivatives of conformal mappings and general facts from analysis of the harmonic measures, such as continuity and non-negativity of the $\beta$-spectrum. This is the first known analytic description of multi-fractal spectrum for unbounded conformal mappings.

The spectrum (\ref{results}) shows two transitions, one of which happens at negative $q=-1-\frac{3\kappa}{8}$ due to singularity of the conformal mappings at the tip of the curve. Another transition happens at positive $q=Q(\kappa)$.

Note that these transitions can already be traced in the simplest example of the $\kappa=0$ evolution: In such a case we have a deterministic growth of the ray on the real line and $\mathcal{F}_-=e^t\frac{w}{(1+w)^2}$. Derivative of this map $\mathcal{F}_-'=e^t\frac{(1-w)}{(1+w)^3}$ vanishes on the tip of the ray at $w=1$, which, according to (\ref{beta_spectrum}), results in transition at $q=-1$. On the other hand the derivative of this map goes to infinity as $w\to -1$ and we have transition at $q=Q(0)=1/3$.

It worth to mention that several points of this spectrum (interior problem for $q=2$ and $\kappa=2$, $\kappa=6$) have been recently found using computer algebra by B.Duplantier et al in \cite{DNNZ}.

\section{Derivative expectations}

To find $\beta^\pm$-spectrum (\ref{beta_spectrum}), one needs to evaluate moments of derivatives of conformal mappings. It turns out that moments of derivatives satisfy linear partial differential equations of the second order. To derive such equations it is convenient to change variable $w\to we^{\mathrm{i} B(t)}$ and define the conformal transformations  $z=f_\pm(w,t)$ in the ``rotating frame'', in which the immovable point $w=1$ on the unit circle is mapped to the moving tip of the curve:
\begin{equation}
f_\pm(w,t)=\mathcal{F}_\pm(we^{\mathrm{i} B(t)},t)=e^{t+\mathrm{i} B(t)}\left (w+\sum_{j=1\mp 1}^\infty f^\pm_j(t)w^{\mp j} \right) .
\label{rotation}
\end{equation}
We remind that the unbounded mapping $f_-$, corresponding to the interior whole-plane SLE$_\kappa$ (\ref{WholeSLE_fixed}), (\ref{T_limit_SLE_int}), sends the unit disc in the $w$-plane to the complement of a curve that starts at some point and goes to infinity in the $z$-plane. On the other hand, the bounded mapping $f_+$, corresponding to the exterior whole-plane SLE$_\kappa$ (\ref{WholeSLE_fixed}), (\ref{T_limit_SLE_ext}), sends the complement of the unit disc in the $w$-plane to the complement of a bounded curve that starts and ends at some finite points in the $z$-plane.

We now define the following functions
\begin{equation}
\rho(w,\bar w| q; \kappa)=e^{-qt}\langle (f'_\pm(w,t) \bar f'_\pm(\bar w,t))^{q/2}\rangle=\lim_{T\to\infty}e^{\mp qT}\langle (F'_\pm(we^{\mathrm{i} \tilde B(T)},T) \bar F'_\pm(\bar we^{-\mathrm{i} \tilde B(T)},T))^{q/2}\rangle,
\label{correlator}
\end{equation}
where the prime denotes derivative wrt $w$ or $\bar w$ (in general, we do not suppose that the second argument $\bar w$ of the above function is a complex conjugate of the first one $w$).

To estimate $\rho$ one can use the following differential equation
\begin{equation}
L[\rho](w,\bar w|q;\kappa)=\sigma q \rho(w,\bar w|q;\kappa),
\label{Lrho}
\end{equation}
$$
L=-\frac{\kappa}{2}\left(w\frac{\partial}{\partial w}-\bar w\frac{\partial}{\partial \bar w}\right)^2+\frac{w+1}{w-1}w\frac{\partial}{\partial w}+\frac{\bar w+1}{\bar w-1}\bar w\frac{\partial}{\partial \bar w}-\frac{q}{(w-1)^2}-\frac{q}{(\bar w-1)^2}+q
$$
where $\sigma=-1$ for the interior problem and $\sigma=1$ for the exterior problem respectively.

Below, we present a simple derivation of equation (\ref{Lrho}) which relies on procedure first introduced by Hastings \cite{H} (for a derivation using formal Ito calculus one can apply e.g. approach presented in \cite{BS}).

We start with the exterior whole-plane SLE$_\kappa$. From (\ref{correlator}) it follows that
\begin{equation}
\frac{\partial}{\partial t}\langle g \rangle=q \langle g \rangle, \quad g=(f'_+(w,t)\bar f'_+(\bar w,t))^{q/2}
\label{t}
\end{equation}
Using the fact that for the Loewner chain (\ref{WholeSLE_fixed}) we have the composition rule $\mathcal{F}_+(w,t+\delta t)=\mathcal{F}_+\left(\delta \mathcal{F}(w,\delta t),t\right)$, where $\delta \mathcal{F}$ itself satisfies the Loewner chain equation
\begin{equation}
\frac{\partial \delta \mathcal{F}(w,\delta t)}{\partial \delta t}=w\frac{\partial \delta \mathcal{F}(w,\delta t)}{\partial w}\frac{w+e^{\mathrm{i} B(t+\delta t)}}{w-e^{\mathrm{i} B (t+\delta t)}}, \quad \delta \mathcal{F}(w,\delta t=0)=w,
\label{delta F}
\end{equation}
in the ``rotating frame" (\ref{rotation}) we get
\begin{equation}
f_+(w, t+\delta t)=f_+\left(e^{-\mathrm{i} B(t)}\delta \mathcal{F}\left(we^{\mathrm{i} B(t+\delta t)},\delta t\right), t\right).
\label{r_composition}
\end{equation}
Solving the differential equation (\ref{delta F}) up to the first order in $\delta t$ (up to the second order in $\delta B=B(t+\delta t)-B(t)$) we obtain
$$
e^{-\mathrm{i} B(t)}\delta \mathcal{F}\left(we^{\mathrm{i} B(t+\delta t)},\delta t\right)=w+\delta \phi(w)+\dots, \quad \delta\phi(w)=w\frac{w+1}{w-1}\delta t+\mathrm{i}w\delta B-\frac{w}{2}(\delta B)^2.
$$
Then, with the help of (\ref{t}), (\ref{r_composition}) we conclude that in the first order in $\delta t$:
$$
g(w, \bar w; t+\delta t)=g_+\left(w+\delta\phi(w), \bar w+\delta \bar \phi(\bar w); t\right)\left(1+\delta\phi'(w)\right)^{q/2}\left(1+\delta\bar\phi'(\bar w)\right)^{q/2}+\dots
$$
Equating expectations of RHS and LHS of the above and taking into account that $<\delta B>=0$, $<(\delta B)^2>=\kappa\delta t$ we find the time derivative of $\langle g\rangle$
$$
\frac{\partial}{\partial t}\langle g \rangle=L\left[\langle g \rangle\right],
$$
where differential operator $\mathcal{L}$ is given in (\ref{Lrho}). Finally, with the help of (\ref{correlator}) and (\ref{t}) we arrive at differential equation (\ref{Lrho}) with $\sigma=1$ for the exterior whole-plane SLE$_\kappa$.

Since, according to (\ref{T_limit_SLE_int}), we defined the interior whole-plane SLE as a reversed Markovian process, equation for the interior whole-plane SLE$_\kappa$ has to be derived in the reversed time. This derivation coincides with that of the exterior case except the factor $e^{qt}$ must replace $e^{-qt}$ in (\ref{correlator}), and so we arrive at equation (\ref{Lrho}) with $\sigma=-1$. \\

\section{Exact Solutions and the $\beta$-spectrum}

In the rest of the article we study the interior problem ($\sigma=-1$) only, i.e. unbounded mapping version of the whole plane SLE$_\kappa$. So we drop all $\pm$ sub/superscripts from our notations.

It is convenient to make the following change of dependent variable $\rho$ that eliminates most singular terms in equation (\ref{Lrho})
\begin{equation}
\rho(w, \bar w| q; \kappa)=\left(\left(1-w\right)\left(1-\bar w\right)\right)^\gamma\Theta(w, \bar w| \gamma; \kappa), \quad q=2\gamma+\frac{1}{2}\kappa\gamma-\frac{1}{2}\kappa\gamma^2,
\label{rho_to_theta}
\end{equation}
where $\gamma$ is given by (\ref{gamma}). Then (\ref{Lrho}) becomes
\begin{eqnarray}\nonumber
\left(-\frac{\kappa}{2}(w-1)(\bar w -1)\left(w\frac{\partial}{\partial w}-\bar w\frac{\partial}{\partial \bar w}\right)^2
+(\kappa\gamma-1)(w-\bar w)\left(w\frac{\partial}{\partial w}-\bar w\frac{\partial}{\partial \bar w}\right) \right. \\
\left. + \left(w\bar w-1\right)\left(w\frac{\partial}{\partial w}
+\bar w\frac{\partial}{\partial \bar w}\right)+\left((\kappa-\kappa\gamma+6)w\bar w+\frac{2\kappa\gamma-\kappa-6}{2}(w+\bar w)\right)\gamma\right)\Theta=0,
\label{Ltheta}
\end{eqnarray}

According to (\ref{rotation}), (\ref{correlator}) and (\ref{rho_to_theta}), function $\Theta(w,\bar w)$ has the following series expansion around $w=0$
\begin{equation}
\Theta(w, \bar w| \gamma; \kappa)=\sum_{i=1}^\infty \sum_{j=1}^\infty \theta_{i,j}(\gamma, \kappa)w^{i-1} \bar w^{j-1}, \quad \theta_{1,1}=1.
\label{whole_int}
\end{equation}

Substituting series expansions (\ref{whole_int}) into equation (\ref{Ltheta}), we get
the ``two-dimensional", four-term recurrence relation for $\theta_{i,j}$:
\begin{equation}
\sum_{k=0}^1\sum_{l=0}^1 C_{i,j}^{l,k}\theta_{i-l,j-k}=0, \quad
\theta_{1,1}^-=1, \quad \theta_{i,j<1}=\theta_{i<1,j}=0
\label{r_whole}
\end{equation}
$$
C_{i,j}^{0,0}=-\frac{\kappa}{2}(i-j)^2-i-j+2 , \quad
C_{i,j}^{1,1}=-\frac{\kappa}{2}(i-j)^2+i+j-\kappa\gamma^2+\kappa\gamma+6\gamma-4, \quad
$$
$$
C_{i,j}^{0,1}=-\frac{\kappa}{2}(i-j+1)^2+(1-\kappa\gamma)(i-j+1)+\kappa\gamma^2-\frac{1}{2}\kappa\gamma-3\gamma, \quad C_{i,j}^{1,0}=C_{j,i}^{0,1} ,
$$
Note, that any element $\theta_{i,j}(\gamma,\kappa)$ can be found in a consecutive manner by expressing $\theta_{i,j}$ as a linear combination of 3 elements: $\theta_{i,j-1}$, $\theta_{i-1,j}$ and $\theta_{i-1,j-1}$: At the first step one finds $\theta_{1,2}$, then $\theta_{1,3}$ etc up to $\theta_{1,l}$. Repeating similar procedure for the next row $\theta_{2,j}$ etc, up to the $n$th row, one gets all $\theta_{1..n, 1..n}$.

It follows from the above described recurrent procedure that once a solution of equation (\ref{Lrho}) analytic at $w=0$ is found for a given $q$, it corresponds to expectation of the $q$-th moment of the absolute value of derivative, since such a solution is unique (up to a constant factor).

To determine $\beta$-spectrum we will use an infinite number of families of exact solutions of equation (\ref{Ltheta}) related to finite-band reductions of matrices of expansion coefficients $\theta_{i,j}$.

\begin{figure}
\begin{center}
\includegraphics[width=10cm]{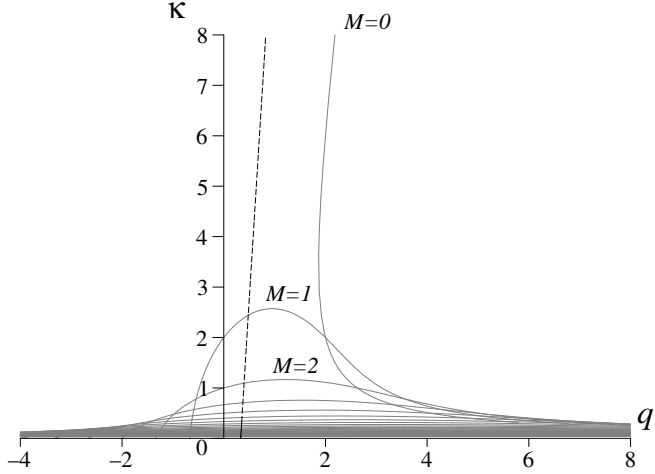}
\caption{Gray curves show infinite families (\ref{kq_int}) in the parametric $(q, \kappa)$-plane along which the $2M+1$-diagonal truncations of solutions take place. Dashed dark curve marks transition in the means $\beta$-spectrum at $q=Q(\kappa)$.}
\end{center}
\end{figure}

In more details: As easy to see from the above described recurrent procedure and relation (\ref{r_whole}), the expansion coefficients $\theta_{i,j}$ vanish for $|i-j|>M$, where $M=0,1,2\dots$, if the recurrence coefficient $C_{i,j}^{0,1}$ vanishes at $j=i+M+1$. As follows from (\ref{r_whole}), such a $2M+1$-band truncation happens when
\begin{equation}
\kappa=2\frac{M+3\gamma}{M^2+2M\gamma+2\gamma^2-\gamma}, \quad q=\frac{\gamma(M+\gamma)(2M+1+\gamma)}{M^2+2M\gamma+2\gamma^2-\gamma}
\label{kq_int}
\end{equation}
The above equations determine infinite number of curves on the parametric $(q, \kappa)$-plane (see Figure 1). In the rest of the paper, with the help of exact solutions, we find means $\beta$-spectrum along these curves and extrapolate it on the whole $(q, \kappa)$-plane.

Note that due to non-negativity of $\kappa$ and according to (\ref{kq_int})
$$
\gamma\ge -\frac{M}{3}.
$$
In the case of the above $2M+1$-diagonal truncations, solutions to (\ref{Ltheta}) can be represented as the Fourier polynomials
\begin{equation}
\Theta(w, \bar w)=\sum_{n=-M}^M w^n f_n\left(w\bar w\right) ,
\label{fourier}
\end{equation}
where coefficients $f_n(\xi), \xi:=w\bar w$ are regular at $\xi=0$. Note, that since $\theta(w, \bar w)$ is real and symmetric, the expansion coefficients are also real and symmetric, i.e. $\bar\theta_{i,j}=\theta_{i,j}$, $\theta_{i,j}=\theta_{i,j}$ and
\begin{equation}
f_{-n}(\xi)=\xi^nf_n(\xi), \quad \bar f_n = f_n.
\label{symmetry}
\end{equation}
Substituting Fourier expansion (\ref{fourier}) into equation (\ref{Lrho}) we get the three-term recurrence relation for $f_n(\xi)$
\begin{equation}
\xi{A}_{n+1} f_{n+1} + {A}_{-n+1} f_{n-1}+\left({B}_n+(1-\xi){C}_n\right) f_{n}+2\xi(\xi-1)\frac{df_n}{d\xi}=0,
\label{rec3}
\end{equation}
where
$$
{A}_n=\frac{\kappa}{2}(n-\gamma)^2+n-3\gamma-\frac{\kappa}{2}\gamma(1-\gamma),
$$
$$
{B}_n=-\kappa (n^2+\gamma^2-\gamma)+6\gamma, \quad {C}_n=\kappa\frac{n^2-2\gamma+2\gamma^2}{2}-n-6\gamma .
$$
Taking into account the symmetry relation (\ref{symmetry}), we see that in the case of the $2M+1$-band truncation, recurrent relation (\ref{rec3}) is a system of $M+1$ first order linear ODEs for $M+1$ functions $f_0(\xi), f_1(\xi),\dots, f_M(\xi)$. By consecutive differentiations and linear transformations, this system can be always ``diagonalized", i.e. decoupled into separate $M+1$th order ODEs for each function $f_i(\xi)$. These equations of $M+1$th order have polynomial in $\xi$ coefficients and their singular points are $\xi=0$, $\xi=1$ and $\xi=\infty$.

As examples, we write down solutions for $M=0$ and $M=1$ cases:

The simplest case is the $M=0$ family of solutions, where we have one-diagonal truncation (i.e. matrix $\theta_{i,j}$ is diagonal) and (\ref{Ltheta}) is reduced to a single first order linear ODE with the following solution
$$
\Theta(w, \bar w)=(1-w\bar w)^{\frac{-3\gamma^2}{2\gamma-1}}, \quad \kappa=\frac{6}{2\gamma-1}, \quad q=\frac{\gamma(\gamma+1)}{2\gamma-1},
$$
where values of $q$ and $\kappa$ are taken from equations (\ref{kq_int}) with $M=0$. According to (\ref{rho_to_theta}), this corresponds to the one-parametric family of solutions $\rho=\rho_0(w,\bar w)$ of fundamental equation (\ref{Lrho}) along the $q=\frac{(2+\kappa)(6+\kappa)}{8\kappa}$ curve in the parametric $(q,\kappa)$-plane
\begin{equation}
\rho_0=\frac{\left((1-w)(1-\bar w)\right)^{\frac{6+\kappa}{2\kappa}}}{\left(1-w \bar w\right)^{\frac{(6+\kappa)^2}{8\kappa}}}, \quad q=\frac{(2+\kappa)(6+\kappa)}{8\kappa} .
\label{int_M_0}
\end{equation}
The second example is the $M=1$ family of solutions (i.e. case of the three-diagonal truncation) $\rho=\rho_1(w,\bar w)$. Here $f_0(\xi)$ and $f_1(\xi)$ each satisfies a second order equation. These equations are amenable to the Gauss hypergeometric type equations and
$$
\rho_1=\left((1-w)(1-{\bar w})\right)^\gamma(1-w\bar w)^{-\frac{(\gamma+1)(3\gamma^2+6\gamma-1)}{2\gamma^2+\gamma+1}}\left(\left(1-\frac{w+\bar w}{2}\right)\Phi_1+\frac{1-3\gamma}{1+\gamma}(1-w\bar w)\frac{w+\bar w}{2}\Phi_2\right)
$$
$$
\Phi_1={}_2F_1\left(\begin{array}{l}\frac{(\gamma+1)(1-3\gamma)}{2\gamma^2+\gamma+1}, \frac{1-\gamma-4\gamma^2}{2\gamma^2+\gamma+1}; \frac{(\gamma+1)^2}{2\gamma^2+\gamma+1} \end{array} \vert w\bar w \right), \quad \Phi_2={}_2F_1\left(\begin{array}{l}\frac{(1-\gamma)(2+\gamma)}{2\gamma^2+\gamma+1}, \frac{2(1-\gamma^2)}{2\gamma^2+\gamma+1}; \frac{3\gamma^2+3\gamma+2}{2\gamma^2+\gamma+1} \end{array} \vert w\bar w \right),
$$
\begin{equation}
\kappa=\frac{2(3\gamma+1)}{2\gamma^2+\gamma+1}, \quad q=\frac{\gamma(\gamma+1)(\gamma+3)}{2\gamma^2+\gamma+1} .
\label{int_M_1}
\end{equation}

The goal of this work is finding the $\beta$-spectra of the whole plane SLE$_\kappa$. For this purpoce we need to look for asymptotic behavior of $2M+1$-band solutions (\ref{fourier}) in vicinity of the unit circle $w\bar w=1$ (i.e. at $\xi\to 1$).

In the case of $2M+1$-diagonal truncation, the system of $2M+1$ linear ODEs (\ref{rec3}) for $f:=\{f_n(\xi), n=-M..M\}$ has $2M+1$ linearly independent solutions that behave at $\xi\to 1$ as
$$
f(\xi) \to (1-\xi)^{-\beta_l}\mathcal{V}_l,
$$
where $\beta_l, l=0..2M$, are $2M+1$ eigenvalues of a $2M+1\times2M+1$ tri-diagonal matrix and $\mathcal{V}_l$ is a $\xi$-independent $2M+1$ component eigenvector corresponding to the eigenvalue $\beta_l$. Indeed, expanding solutions of (\ref{rec3}) at $\xi\to 1$ as follows
\begin{equation}
f_n(\xi)=(1-\xi)^{-\tilde \beta}\psi_n +\dots, \quad n=-M..M
\label{xi_to_1}
\end{equation}
and substituting the above expansion into (\ref{rec3}) we get the following three-term recurrence relation for $\psi_n$
\begin{equation}
R[\psi]=\tilde \beta \psi, \quad R[\psi]_n:=\frac{1}{2}\left({A}_{n+1} \psi_{n+1} + {A}_{-n+1} \psi_{n-1}+{B}_n \psi_{n}\right)
\label{three_diagonal}
\end{equation}
with $\tilde\beta$ being an eigenvalue of the $2M+1\times 2M+1$ three-diagonal matrix $R$.

It is convenient to introduce the following function
\begin{equation}
\Psi(\varphi):=\sum_{n=-M}^M\psi_n e^{\i n \varphi}, \quad \bar\Psi=\Psi
\label{psi}
\end{equation}
As follows from (\ref{fourier}), (\ref{xi_to_1}) this is the ``angular part" of $\Theta(w,\bar w)$ at the unit circle $w=e^{\i\varphi}$, i.e.
\begin{equation}
\Psi\left(\varphi\right)=\lim_{r\to 1-}\frac{\Theta\left(re^{\i\varphi}, re^{-\i\varphi}\right)}{(1-r^2)^{\tilde\beta}}
\label{Psi_angular}
\end{equation}
To relate the above eigenvalue to the $\beta$-spectrum, we note that according to definitions (\ref{beta_spectrum}) and (\ref{correlator})
$$
\beta=\lim_{\epsilon \to 0+}\frac{\log \int_0^{2\pi}\rho\left(e^{-\epsilon+\i\varphi},e^{-\epsilon-\i\varphi}|q;\kappa\right)d\varphi}{-\log \epsilon},
$$
Then, with help of (\ref{rho_to_theta}), and the fact that function $\left((1-w)(1-\bar w)\right)^\gamma$ ceases to be integrable on the unit circle $w=e^{\i \varphi}$ when $\gamma<-1/2$, from the previous equation and eq. (\ref{Psi_angular}) we have
\begin{equation}
\beta=\left\{\begin{array}{ll} \tilde\beta-2\gamma-1, & \gamma\le-\frac{1}{2} \\ \tilde\beta, & \gamma\ge-\frac{1}{2}\end{array}\right., \quad \tilde\beta \in \{\beta_l, l=0..2M\},
\label{tilde_beta}
\end{equation}
provided $\Psi(\varphi)$ does not have zeros at $\varphi=0$ when $\gamma<-1/2$ (we will see later that this is the case).
Therefore, the value of the means $\beta$-spectrum of the whole-plane SLE$_\kappa$ is determined by an eigenvalue of the three diagonal matrix (\ref{three_diagonal}).

The eigenvalues $\{\beta_l, l=0..2M\}$ can be found exactly: the spectrum $\beta_l=\beta_l(M,\gamma)$ is a quadratic function of $l$ with the biggest eigenvalue being either $\beta_0$ or $\beta_{2M}$ depending on values $M$ and $\gamma$. Below we will show that $\tilde \beta=\max(\beta_l, l=0..2M)$, i.e., in the case of $2M+1$ truncation, the $\beta$-spectrum is determined by the biggest eigenvalue of the three-diagonal matrix (\ref{three_diagonal}).

As a consequence one can observe transition in the means $\beta$-spectrum at points of the parametric $(q,\kappa)$-plane where $\beta_0=\beta_{2M}$. Such a transition happens when $q=Q(\kappa)$.

It is interesting to note that, in an addition to $\kappa=0$ example given in the introduction, another simple explicit example of such a transition can be observed on exact solution (\ref{int_M_1}) for the $M=1$ family, where at $|w|\to 1$, $f_0\to {\rm Const}\left(1-w\bar w\right)^{-\frac{(\gamma+1)(3\gamma^2+6\gamma-1)}{2\gamma^2+\gamma+1}}$ when $\gamma>\frac{\sqrt{57}-5}{16}$ and $f_0\to {\rm Const}\left(1-w\bar w\right)^{-\frac{\gamma^2(3\gamma+1)}{2\gamma^2+\gamma+1}}$ when $-1/3<\gamma<\frac{\sqrt{57}-5}{16}$.

\section{Computation of the spectrum}

To get the $\beta$-spectrum we need to determine eigenvalues of the three-diagonal matrix (\ref{three_diagonal}). Due to mirror symmetry (i.e. symmetry w.r.t. reflections $n \to -n$) its eigenvectors are either symmetric or antisymmetric in $n$. By reality and symmetry of $\rho$ and $\Theta$ we need to consider only symmetric eigenvectors
$$
\psi_n=\psi_{-n}, \quad \Psi(\varphi)=\Psi(-\varphi)
$$
It is now convenient to rewrite the three-term relation (\ref{three_diagonal}) in the form of the second order linear ODE for $\Psi(\varphi)$:
\begin{equation}
\frac{\kappa}{2}\left(1-\cos(\varphi)\right)\Psi''(\varphi)-(1-\kappa\gamma)\sin(\varphi)\Psi'(\varphi)+
\left(\left(\kappa\frac{2\gamma-1}{2}-3\right)\gamma\cos(\varphi)-\left(\kappa\frac{\gamma-1}{2}-3\right)\gamma-\tilde\beta\right)\Psi(\varphi)=0
\label{LPsi}
\end{equation}
It is also convenient to make the change of the independent variable $x=\frac{1-\cos(\varphi)}{2}$ in the above equation and also reparametrize $\tilde\beta=\tilde\beta(\lambda)$ as
\begin{equation}
\tilde\beta=3q-1-\frac{q}{M+3\gamma}+\left(\kappa\frac{1-\lambda-2M-12\gamma}{4}+1\right)(2M-\lambda)
\label{beta_lambda}
\end{equation}
After such changes, equation (\ref{LPsi}) is amenable to the hypergeometric form and has the following general solution
$$
\Psi(\varphi)=\mathcal{C}_1x^{\lambda/2}g_1(x)+\mathcal{C}_2x^{(\lambda-1)/2}g_2(x)
$$
with
\begin{equation}
g_1(x)={}_2F_1\left(a,b; \frac{1}{2}+a+b | x\right), \quad
g_2(x)=\sqrt{x(1-x)}\,\,{}_2F_1\left(\frac{1}{2}+a,\frac{1}{2}+b; \frac{3}{2} |1-x \right)
\label{g_1g_2}
\end{equation}
where
$$
a=\frac{\lambda}{2}+\gamma-\frac{1}{\kappa}-\frac{\sqrt{1+2q\kappa}}{\kappa}, \quad
b=\frac{\lambda}{2}+\gamma-\frac{1}{\kappa}+\frac{\sqrt{1+2q\kappa}}{\kappa} ,
$$
or in terms of $\gamma$ and $M$ (see Eq.(\ref{kq_int}) )
\begin{equation}
a=\frac{\lambda}{2}-M, \quad
b=\frac{\lambda}{2}+\gamma\frac{3M+1+4\gamma}{M+3\gamma}
\label{ab_M}
\end{equation}
As seen from (\ref{g_1g_2}) and (\ref{ab_M}) $g_1$ has a form of the Fourier polynomial (\ref{psi}) which is an even function of $\varphi$ if $\lambda$ is an even non-negative integer. On the other hand when $\lambda$ is an odd positive integer, $g_2$ is a Fourier polynomial which is odd in $\varphi$. Therefore, in the case of $2M+1$-diagonal reduction we have to choose an eigenvalue from the $\lambda=l \in \{0,2,4,\dots 2M\}$ subset of (\ref{beta_lambda}) with eigenfunctions which (up to a constant factor) equal
\begin{equation}
\Psi(\varphi) \in x^{l/2}{}_2F_1\left(\frac{\l}{2}-M, \frac{l}{2}+\gamma\frac{3M+1+4\gamma}{M+3\gamma}; \frac{1}{2}+l-M+\gamma\frac{3M+1+4\gamma}{M+3\gamma}\left|x\right.\right), \quad l=0,2,4,\dots 2M
\label{psi_l}
\end{equation}
Corresponding eigenvalues equal
\begin{equation}
\beta_l=\frac{2(M+3\gamma)\gamma^2-(2M^2+M-8\gamma^2+\gamma)l+(M+3\gamma)l^2}{2(M^2+2M\gamma+2\gamma^2-\gamma)}
\label{beta_l}
\end{equation}
The biggest eigenvalue, depending on $M$ and $\gamma$, is either
\begin{equation}
\beta_0=\frac{M+3\gamma}{M^2+2M\gamma+2\gamma^2-\gamma}\gamma^2=\frac{\kappa\gamma^2}{2}
\label{beta_0}
\end{equation}
or
\begin{equation}
\beta_{2M}=\frac{(M+\gamma)(6M\gamma-M+3\gamma^2)}{M^2+2M\gamma+2\gamma^2-\gamma}=3q-\frac{1}{2}-\frac{1}{2}\sqrt{1+2q\kappa}
\label{beta_2M}
\end{equation}
These two eigenvalues are equal at the transition point
\begin{equation}
\gamma=\gamma_M:=\frac{\sqrt{36M^2+20M+1}-6M+1}{16}
\label{at}
\end{equation}
or equivalently at $q=Q(\kappa)$.

Now we will show that in the case of the $2M+1$-diagonal reduction $\tilde\beta$ equals the biggest eigenvalue.

Indeed, by non-negativity of derivative moments, $\Psi(\varphi)$ should be non-negative and cannot oscillate on the unit circle. In the case when $\gamma>\gamma_M$, (i.e. in the case when $\beta_{2M}>\beta_0$) the only eigenfunction that does not oscillate on the unit circle is that corresponding to the biggest eigenvalue $\beta_{2M}$. According to (\ref{psi_l}), up to a constant factor, it equals
\begin{equation}
\Psi=(1-\cos(\varphi))^M
\label{Psi_M}
\end{equation}
Instead of direct proving that this is the only non-oscillating eigenfunction when $\gamma>\gamma_M$, we note that, as follows from (\ref{beta_l}), the only positive even eigenvalues in the vicinity of point $\gamma=\gamma_M$ are $\beta_0$ and $\beta_{2M}$. The eigenfunction that corresponds to $l=0$ in (\ref{psi_l}) equals to $1$ at $x=0$ and is negative at $x=1$, when $\gamma>\gamma_M$. Therefore the only choice for $\tilde\beta$ in some vicinity of $\gamma=\gamma_M$ for $\gamma>\gamma_M$ is $\tilde\beta=\beta_{2M}$. Also, for all $\gamma>\gamma_M$ the rest of eigenvalues $\beta_{2i}(\gamma), i=0..M-1$ is strictly less than $\beta_{2M}$. Therefore by continuity of the $\beta$-spectrum, on the right from transition point $\gamma=\gamma_M$ we have
$$
\tilde\beta=\beta_{2M} \quad {\rm if} \quad \gamma\ge\gamma_M .
$$
On the left from the transition point $\gamma=\gamma_M$, the bigger eigenvalue is $\beta_0$. Here, not only the eigenfunction (\ref{Psi_M}) is non-negative. To proceed, we note that, as follows from (\ref{beta_l}), all the $\beta_{2i}$ except $\beta_0$ are negative when $\gamma<\sqrt{M^2+\frac{M}{3}}-M$. Therefore, in the range $-\frac{M}{3}\le\gamma\le\sqrt{M^2+\frac{M}{3}}-M$, $\tilde\beta=\beta_0$. On the rest of the interval on the left of $\gamma=\gamma_M$, i.e. at $\sqrt{M^2+\frac{M}{3}}-M<\gamma<\gamma_M$, the only positive eigenvalues are $\beta_0$ and $\beta_{2M}$, but since $\beta_0>\beta_{2M}$ for $\gamma<\gamma_M$, by continuity of the beta spectrum we have
$$
\tilde\beta=\beta_0 \quad {\rm if} \quad -\frac{M}{3}\le \gamma \le \gamma_M .
$$
Therefore, in the case of the $2M+1$-diagonal families of solutions $\beta$-spectrum corresponds to biggest eigenvalue of the three-diagonal matrix
(\ref{three_diagonal}) and according to (\ref{beta_0}), (\ref{beta_2M}), (\ref{at}) along the infinite number of curves (\ref{kq_int}) in the parametric $(q,\kappa)$-plane we have
$$
\tilde\beta=\left\{\begin{array}{ll}
\frac{1}{2}\kappa\gamma^2, & q \le Q(\kappa) \\
3q-\frac{1}{2}-\frac{1}{2}\sqrt{1+2q\kappa}, & \quad  q \ge Q(\kappa)
\end{array}\right.
$$
which, together with (\ref{tilde_beta}), leads to the conclusion that along these curves, the integral means $\beta$-spectrum of the whole-plane SLE$_\kappa$ is given by equation (\ref{results}). Therefore, we have derived the main result of the paper (i.e. equation (\ref{results})) for $(q, \kappa)$ belonging to these curves (see Figure 1).

Finally, we put forward the hypothesis that this result can be continued from the infinite family of curves to the upper $\kappa\ge 0$ parametric $(q, \kappa)$-plane.

\section{Concluding Remarks}

In summary: By elementary arguments, in the case of interior problem, any regular at $w=0$ solution of the fundamental equation (\ref{Lrho}) is unique (up to constant factor) and therefore provides a point in the $\beta$-spectrum. Continuing analytically $\beta(q, \kappa)$ spectrum derived along the infinite number of curves in the parametric $(q, \kappa)$-plane, we have presented exact description of the multi-fractal spectrum for the unbounded whole-plane SLE$_\kappa$.

The question of proof of continuation of the spectrum from the curves to the plane remains open. However, it is worth to mention that arguments leading to construction of function $\Psi(\varphi)$ can be formally applied to the general case (i.e. not only for the $2M+1$-diagonal solutions), which leads to equation for $\Psi$ of the form (\ref{LPsi}) with $\Psi$ being, in general, infinite Fourier series. Repeating the above computations for infinite series one can find that $\Psi(\varphi)$ corresponding to the $\beta$-spectrum (\ref{results}) is non-negative also in the generic case.

Also note that the $\beta$-spectrum at $q<Q(\kappa)$ coincides with that predicted from the quantum gravity by B. Duplantier \cite{D}.

Another interesting question is study of the Loewner evolution driven by non-Brownian Levy processes. In difference from the case of the Schramm-Loewner Evolution the moments of derivatives for such processes satisfy fundamental equations which are not differential. However, one can rewrite such equations in the form of a nine-term ``two-dimensional" recurrence relation for the Taylor coefficients of function $\rho(w,\bar w)$ which also admits some finite-diagonal reductions \cite{L}, leading to exact solutions for some Levy processes. It will be interesting to study such solutions.

\vspace{5mm}

\noindent
{\large\textbf{Acknowledgement}}\\

We would like to acknowledge help received from V.~Spiridonov and A.~Zhedanov.
The work of authors has been supported by the European Commission 7th framework IEF grants.

\end{document}